# Estimating States for Nonlinear Systems Using the Particle Filter

**Johnny Condori [a,\*], Amin Maghareh [a], Shirley Dyke [a,b]**

[a] Lyles School of Civil Engineering, Purdue University, West Lafayette, Indiana, USA
[b] School of Mechanical Engineering, Purdue University, West Lafayette, Indiana, USA

**ABSTRACT**

Kalman filtering has been traditionally applied in three application areas of estimation, state estimation, parameter estimation (a.k.a. model updating), and dual estimation. However, Kalman filter is often not sufficient when experimenting with highly uncertain nonlinear dynamic systems. In this study, a nonlinear estimator is developed by adopting a particle filter algorithm that takes advantage of measured signals. This approach is shown to significantly improve the ability to estimate states. To illustrate this approach, a model for a nonlinear device coupled with a hydraulic actuator plays the role of an actual plant and a nonlinear algebraic function is considered as an approximation of the nonlinear device, thus generating non-parametric and parametric uncertainties. Then we use displacement and force signals to improve the distribution of the states by resampling the set of particles. Finally, all of the states are estimated from these posterior density functions. A set of simulations considering three different noise levels demonstrates that the performance of the particle filter approach is superior to the Kalman filter, yielding substantially better performance when estimating nonlinear physical systems in the presence of modeling uncertainties.

## 1. Introduction

The mathematical models of most physical systems can be captured with a set of differential equations (a.k.a., a dynamic system). A large number of engineering problems is encompassed in this set if we restrict our attention to the case in which the state is the solution of a system of ordinary differential equations [1]. Such a representation usually arises quite naturally from the application of fundamental physics principles known to govern the particular dynamic system's behavior [2]. The input parameters to a dynamic model include physical constants, equations of states, boundary and initial conditions and so on. Accurate knowledge of these parameters and variables is often impossible, either due to measurement limitations and/or the random nature of the physical system. We thus treat the input parameters as well as the dynamic states as random variables, expressed as uncertainty [3]. In certain applications (e.g., tracking control, nonlinear estimation, and fault diagnosis), it is essential to explicitly address parametric and non-parametric modeling uncertainties, and there has been a growing interest in the application of stochastic modeling of dynamic systems, uncertainty quantification and estimation [1,3–5].

      The subject of estimation is an important topic within system dynamics and control theory. Estimation is extremely useful in a diverse range of real world problems including state estimation, parameter estimation and dual estimation [6]. Figure 1 depicts the hierarchy of Bayesian estimation techniques. Here, Kalman filters are the most popular approaches [7]. These filters are relatively easy to design, and they often provide good estimation results. However, Kalman filters can lead to poor estimation for several reasons, such as nonlinearities, non-parametric modeling uncertainties, and ill-conditioning of the covariance matrix [8]. The fact is that almost all physical systems are nonlinear in nature although the operation of these systems may be described by linear dynamic systems. Such a model is reasonable only if their mode of operation does not deviate excessively from the chosen nominal conditions [5,8]. For instance, in control theory the first step in designing a control system for a certain physical plant is to

---
[\*] Corresponding author
E-mail address: jcondori@purdue.edu (J. Condori)

identify a meaningful dynamic model of the plant which captures its essential dynamics within an operational range of interest [1]. For nonlinear dynamic systems, robust nonlinear control techniques have been successfully applied in a variety of practical problems [9–12]. These control strategies are designed based on the stochastic modeling of the plant and generally require full state measurement [1,13]. Thus, accurate real-time state estimation of the plant becomes critical in this class of nonlinear control problems.

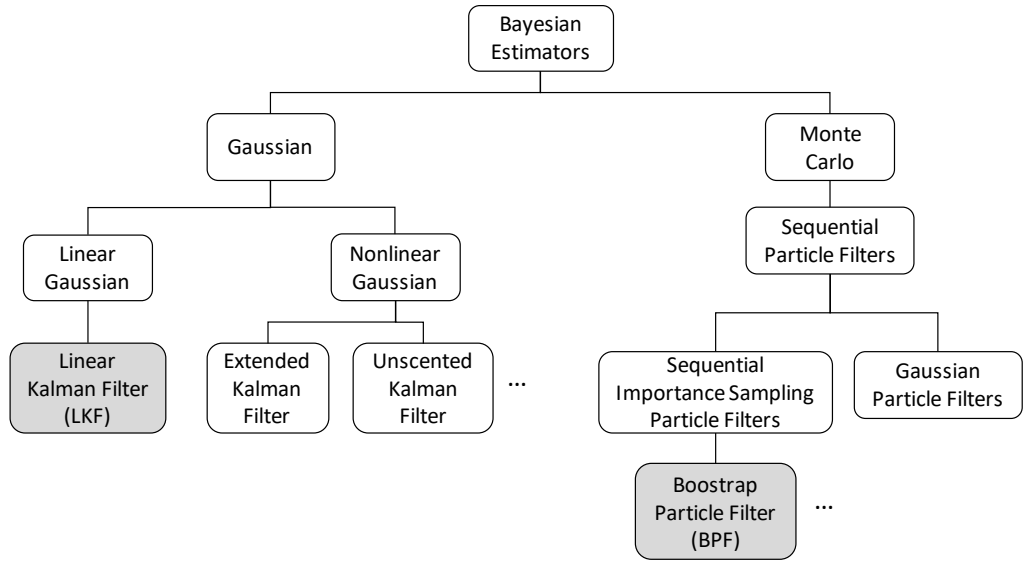

**Figure 1**. Hierarchy of Bayesian estimators [14].

This paper focuses on the comparison of two state estimation techniques for a nonlinear dynamic system, including the *Kalman filter* and the *particle filter*. The approach used for nonlinear estimation with each of these two techniques is provided, with an emphasis on the Bayesian approach used in the particle filter. In a numerical case study, the plant, which is a nonlinear physical specimen coupled with a hydraulic actuator system, is transformed into a controllable canonical representation to generate a process model. Modeling errors are considered by using an algebraic representation of the true nonlinear system dynamics in the estimator. The estimation results demonstrate that the performance of the particle filter aproach is superior to the Kalman filter for this nonlinear system. Since full state measurement is essential in implementing nonlinear robust control of the plant, this improvement is expected to lead to enhanced robust control of the nonlinear plant in real-time scenarios.

## 2. Methodology

Consider a nonlinear dynamic system governed by the equations

$$\dot{\mathbf{x}} = f(\mathbf{x}, \mathbf{u}, \mathbf{w}) \qquad (1)$$

$$\mathbf{y} = h(\mathbf{x}, \mathbf{u}, \mathbf{v}) \qquad (2)$$

where $\mathbf{x}$ is the vector of state variables, $\dot{\mathbf{x}}$ is a vector of their derivatives with respect to time, $\mathbf{u}$ is the vector representing the inputs to the system, $\mathbf{y}$ is the output vector containing variables of interest (including measured signals), and $\mathbf{w}$ and $\mathbf{v}$ are process and measurement noise, respectively. Figure 2 shows a block diagram with the input-output relationship for this dynamic system.

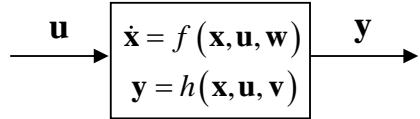

**Figure 2**: Block diagram of nonlinear dynamic system.

The objective is to design an estimator that uses measured information to produce an estimate of the state vector, denoted $\hat{\mathbf{x}}$. Two approaches are considered here for estimating the states, the *Kalman filter* and the *bootstrap particle filter*.

**Kalman Filter**

The Kalman filter is a widely used and well-known approach that, in the case of a nonlinear dynamic system, requires linearization of the system [15]. Thus, the form of the linearized stochastic difference equations for Eqs. 1 and 2 are

$$\mathbf{x}_{k+1} = \mathbf{A}\mathbf{x}_k + \mathbf{B}\mathbf{u}_k + \mathbf{w}_k \qquad (3)$$

$$\mathbf{y}_k = \mathbf{C}\mathbf{x}_k + \mathbf{v}_k \qquad (4)$$

where the process noise, $\mathbf{w}$, and the measurement noise, $\mathbf{v}$, are assumed to be independent, white, and normally distributed signals with probability density functions $p(\mathbf{w}) \sim N(0, Q)$ and $p(\mathbf{v}) \sim N(0, R)$, respectively. $\mathbf{A}$ is the system matrix, $\mathbf{B}$ is the input matrix, and $\mathbf{C}$ is the output matrix.

The Kalman filter propagates and updates the mean and covariance of the distribution. In this approach, the distribution of the states of the system is assumed to remain Gaussian at every iteration. Table 1 provides the algorithm, and the details of this method are widely available in the literature [16,17].

**Table 1**. Procedure for the Kalman filter.

| Step | Equation |
|---|---|
| 1. **Initialization**: Initial estimates for: | $\hat{\mathbf{x}}_{k-1}$ and $P_{k-1}$ |
| 2. **Prediction** (time update): | |
|    a. Project the state ahead | $\hat{\mathbf{x}}_k^- = \mathbf{A}\hat{\mathbf{x}}_{k-1} + \mathbf{B}\mathbf{u}_{k-1}$ |
|    b. Project the error covariance ahead | $P_k^- = \mathbf{A}P_{k-1}\mathbf{A}^T + Q$ |
| 3. **Correction** (Measurement Update): | |
|    a. Compute the Kalman gain | $K_k = P_k^- H^T (HP_k^- H^T + R)^{-1}$ |
|    b. Update estimate using measurement $\mathbf{y}_k$ | $\hat{\mathbf{x}}_k = \hat{\mathbf{x}}_k^- + K_k(\mathbf{y}_k - H\hat{\mathbf{x}}_k^-)$ |
|    c. Update the error covariance | $P_k = (I - K_h H)P_k^-$ |

## Particle Filter

An alternate method that is directly applicable to state estimation of nonlinear systems is the particle filter. The particle filter is a Bayesian approach based on recursive inference, and describes the current probability density distributions of the states conditioned on prior observations, as explained in Table 2. The discussion is restricted to Markovian, nonlinear, non-Gaussian models. For a stochastic system, the distribution describing the states of the system at time $k$ is given by [14,18]

$$\begin{aligned} &p(\mathbf{x}_0) \\ &p(\mathbf{x}_k | \mathbf{x}_{k-1}) \quad , k \geq 1 \quad \text{(transition equation)} \\ &p(\mathbf{y}_k | \mathbf{x}_k) \quad , k \geq 1 \quad \text{(emission equation, marginal distribution)} \end{aligned} \quad (5)$$

where

$\mathbf{x}_k$: vector of unobservable states modeled as a Markov process with initial distribution $p(\mathbf{x}_0)$.
$\mathbf{y}_k$: vector of observations assumed to be conditionally independent given the process $\mathbf{x}_k$.

At time $k$, the states of the system are $\mathbf{x}_{0:k} \triangleq \{\mathbf{x}_0,...,\mathbf{x}_k\}$ and the observations are denoted as $\mathbf{y}_{1:k} \triangleq \{\mathbf{y}_1,...,\mathbf{y}_k\}$. The goal is to estimate the posterior distribution $p(\mathbf{x}_{0:k}|\mathbf{y}_{1:k})$ using Bayes' theorem by relating this posterior distribution to the likelihood $p(\mathbf{y}_{1:k}|\mathbf{x}_{0:k})$ and the prior distribution $p(\mathbf{x}_{0:k})$ as

$$p(\mathbf{x}_{0:k}|\mathbf{y}_{1:k}) = \frac{p(\mathbf{y}_{1:k}|\mathbf{x}_{0:k})p(\mathbf{x}_{0:k})}{p(\mathbf{y}_{1:k}) = \int p(\mathbf{y}_{1:k}|\mathbf{x}_{0:k})p(\mathbf{x}_{0:k})d\mathbf{x}_{0:k}}. \quad (6)$$

The posterior distribution and the likelihood define the statistical model for estimation. Thus, the solution to the estimation problem is defined by Eqs. 5 and 6. Recursively, the idea is to update the *prior density* $p(\mathbf{x}_{0:k})$ to obtain the *posterior density* $p(\mathbf{x}_{0:k}|\mathbf{y}_{1:k})$ by computing the likelihood $p(\mathbf{y}_{1:k}|\mathbf{x}_{0:k})$ using observed data $\mathbf{y}_{1:k}$.

Gordon *et al*. [7] proposed a recursive algorithm to implement the particle filter called the *bootstrap filter*. This algorithm is based on the concept of bootstrap resampling [19,20]. In this algorithm, the state vector $\mathbf{x}_{k+1}$ evolves using a discrete version of the process model, and using the observed data (measurements) $\mathbf{y}_k$ as

$$\mathbf{x}_{k+1} = f_k(\mathbf{x}_k, \mathbf{w}_k) \quad (7)$$

$$\mathbf{y}_k = h_k(\mathbf{x}_k, \mathbf{v}_k) \quad (8)$$

where $f_k$ is the system transition function, $\mathbf{w}_k$ is a zero-mean, white-noise discrete signal that is independent of the states, $h_k$ is the measurement function, and $v_k$ is a zero-mean, white-noise discrete signal that is independent of both the states and the process noise $\mathbf{v}_k$.

Table 2. Procedure for the bootstrap particle filter algorithm [7].

| Step | Equation |
|---|---|
| 1. **Initialization**: Initialize the posterior density (k=0) by generating $N$ particles $\{\mathbf{x}_0(i) : i = 1, ..., N\}$ and selecting an initial distribution, mean, and variance for $p(\mathbf{x}_0)$. | $p(\mathbf{x}_0 \mid \mathbf{y}_o) = p(\mathbf{x}_0)$ |
| 2. **Prediction:** Propagate each particle to the next ($k$th) time step to compute the prior density $p(\mathbf{x}_k \mid \mathbf{y}_{1:k-1})$ via the process equation, Eq. 7, to obtain samples $\{\mathbf{x}_k^*(i) : i = 1, ..., N\}$ | $p(\mathbf{x}_k \mid \mathbf{y}_{1:k-1}) = \int p(\mathbf{x}_k \mid \mathbf{x}_{k-1}) p(\mathbf{x}_{k-1} \mid \mathbf{y}_{1:k-1}) d\mathbf{x}_{k-1}$ |
| 3. **Update:** Find the posterior density $p(\mathbf{x}_k \mid \mathbf{y}_{1:k})$ | $p(\mathbf{x}_k \mid \mathbf{y}_{1:k}) = \dfrac{p(\mathbf{y}_k \mid \mathbf{x}_k) p(\mathbf{x}_k \mid \mathbf{y}_{1:k-1})}{\int p(\mathbf{y}_k \mid \mathbf{x}_k) p(\mathbf{x}_k \mid \mathbf{y}_{1:k-1}) d\mathbf{x}_k}$ |
| a. Compute the likelihood of each particle using the current observation from the measurement equation Eq. 8. | $p(\mathbf{y}_k \mid \mathbf{x}_k^*(i))$ |
| b. Normalize the weights (for each particle) and scale the respective likelihood. At this point, a set of samples $\{\mathbf{x}_k^*(i) : i = 1, ..., N\}$ each with probability mass $q_i$ are generated. | $q_i = \dfrac{p(\mathbf{y}_k \mid \mathbf{x}_k^*(i))}{\sum_{j=1}^{N} p(\mathbf{y}_k \mid \mathbf{x}_k^*(j))}$ |
| c. Resample each of the particles in the set $\{\mathbf{x}_k^*(i) : i = 1, ..., N\}$ to obtain $\{\mathbf{x}_k(i) : i = 1, ..., N\}$. The resampled particles converge to the required posterior density $p(\mathbf{x}_k \mid \mathbf{y}_{1:k})$ as the number of particles $N$ tends to infinity. | Draw a sample randomly from a uniform distribution (0,1) so that the sample $\mathbf{x}_k^*(M)$ that corresponds to $\sum_{j=0}^{M-1} q_j < u_i \leq \sum_{j=0}^{M} q_j$ is chosen to be the sample $\mathbf{x}_k(i)$ with an approximate posterior density. Repeat this process for each particle in the set $\{\mathbf{x}_k^*(i) : i = 1, ..., N\}$ |
| d. Compute necessary statistics such as the mean. | |

# 3. Numerical Case Study

To validate the approach and demonstrate the performance of this nonlinear estimator, both methods are applied to estimate the states of a realistic nonlinear system. Numerical results of the state estimates obtained using the particle filter are compared to those obtained with the Kalman filter. The plant in this case study includes a nonlinear physical specimen coupled with a servo-controlled hydraulic actuator. As an application, together these could represent the physical component of a Real-time Hybrid Simulation test [21–24], i.e., the transfer system plus the experimental substructure of a partitioned structural system.

**Controllable Canonical Representation of the Plant**

In this nonlinear plant (see Fig. 3), the principal components are the servo-valve, hydraulic actuator, analog controller, and physical specimen. This plant takes a single command input, $u$, and generates two outputs, the actuator force, $F$, and the actuator displacement, $x$.

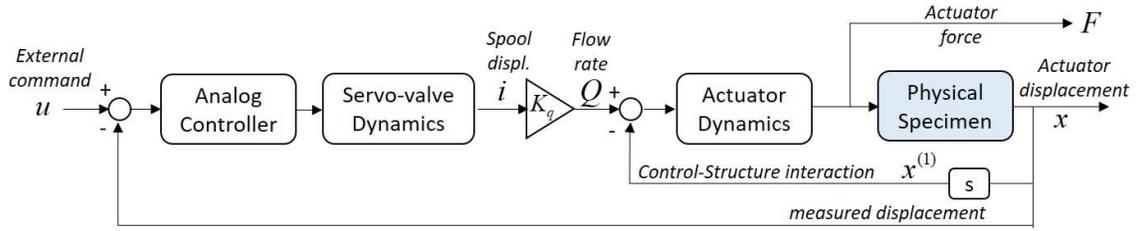

**Figure 3**. Block diagram showing the components of the plant

Maghareh et al. [25,26] showed that this plant can be represented in the controllable canonical form

$$x^{(n)} = f(\mathbf{x}, F) + bu \tag{9}$$

where $x^{(n)}$ denotes the $n^{\text{th}}$ time derivative of the actuator displacement, and $\mathbf{x}$ denotes the vector of the nonlinear states. In this model $f(\mathbf{x}, F)$ is given by

$$f(\mathbf{x}, F) = C_1 x + C_2 x^{(1)} + C_3 x^{(2)} + C_4 x^{(3)} + C_5 F + C_n \tag{10}$$

where $F$ is governed by a nonlinear equation representing the physical specimen given by

$$F = mx^{(2)} + h(x, x^{(1)}) \tag{11}$$

and

$$\begin{aligned}
C_1 &= -a_1 \beta_0 / m \\
C_2 &= \beta_1 a_2 / m + \beta_1 (\partial h / \partial x) + a_3 (\partial h / \partial x) \\
C_3 &= \beta_1 (\partial h / \partial x^{(1)}) + a_3 (\partial h / \partial x^{(1)}) - a_2 / m \\
C_4 &= -\beta_1 - a_3 \\
C_5 &= -\beta_1 a_3 / m \\
C_n &= \ddot{h} = \frac{\partial^2 h}{\partial x_1^2} \left(x^{(1)}\right)^2 + \frac{\partial^2 h}{\partial x_2^2} \left(x^{(2)}\right)^2 + 2 \frac{\partial^2 h}{\partial x_1 \partial x_2} x^{(1)} x^{(2)} + \frac{\partial h}{\partial x_1} x^{(2)} + \frac{\partial h}{\partial x_2} x^{(3)}
\end{aligned} \tag{12}$$

**Actual Plant (True Plant)**

Here we use the term *actual plant model* to represent the true physical plant dynamics. In this case study, we consider uncertainty only in the physical specimen. Thus, the parameters associated with the servo-valve, hydraulic actuator, proportional controller, and their dynamics, are treated here as known. The actual physical specimen is represented as follows

$$h_{act}(x, x^{(1)}) = -\frac{c}{m} x^{(1)} - \frac{k}{m} x - \frac{k_n}{m} \arctan(\lambda x). \tag{13}$$

**Nonlinear Nominal Plant**

The term *nominal plant model* is used to represent an approximation of the actual plant used to study the ability of the two estimators to deal with unmodeled dynamics. In the *nonlinear* nominal plant, the dynamics of the physical specimen are assumed as

$$h_{nom}(x, x^{(1)}) = -\frac{c}{m} x^{(1)} - \frac{k}{m} x - \frac{1}{m} \frac{k_n \cdot x\lambda}{\sqrt{1+(x\lambda)^2}}. \tag{14}$$

Note that this expression has similar behavior to that of the actual physical specimen, but there are significant modeling errors present, as there would be in a real experiment.

The parameters of the transfer system (servo-valve, hydraulic actuator, and proportional controller) used in this study were identified in [26]. The parameters of the nonlinear physical device considered in this numerical study are also shown in Table 3.

**Table 3**: Parameters of the transfer system and physical device

| Parameters of the physical device | | | Identified parameters of the transfer system | | |
|---|---|---|---|---|---|
| Parameter | Value | Unit | Parameter | Value | Unit |
| $m$ | 3.8 | kg | $\beta_1$ | 267 | 1/s |
| $c$ | 10 | N·s/m | $a_1\beta_1$ | 2.412 x $10^9$ | mPa/s$^2$ |
| $k$ | 1500 | N/m | $a_2$ | 7.881 x $10^5$ | mPa |
| $k_n$ (actual plant) | 800 | - | $a_3$ | 16.118 | 1/s |
| $k_n$ (nonlinear nominal plant) | 1100 | - | | | |
| $\lambda$ (actual plant) | 250 | - | | | |
| $\lambda$ (nonlinear nominal plant) | 250 | - | | | |

**Linear Nominal Plant**

The *linear nominal plant model* is developed based on data obtained through a simulated dynamic identification of the actual nonlinear plant (i.e., the inherent nonlinear behavior of the actual plant is linearized). This mimics the approach that would typically be used for a linearized control design as done in [23,27–29]. In this approach, the determination of input-output behavior through transfer functions provides a linearized model that contributes to the understanding of the system dynamics in the frequency domain [30]. Three transfer functions are obtained through simulations. These transfer functions are obtained using three band-limited white noise (BLWN) input signals, each with a Gaussian distribution and a bandwidth of 40 Hz. The three BLWN inputs have increasing amplitudes, root mean square (RMS) values of 15.8 mm, 31.6 mm, and 63.3 mm, which correspond to increasing plant displacement responses (here, RMS values of 5 mm, 10 mm, and 20 mm, respectively). Replicating the typical procedure used in the laboratory, an average of these transfer functions (see Figure 4) is obtained to reasonably represent the range of nonlinear behaviors of the actual plant. Transfer functions from the input to the three outputs are obtained: actuator displacement (m), velocity (m/s), and acceleration (m/s$^2$). Notice in Fig. 3 that the servo-valve and hydraulic actuator models are each described by a first-order system. Also, a linearized model of the physical specimen would be a second-order mass-spring-damper (i.e., an oscillator). Thus, the entire linear nominal plant is realized as a fourth-order dynamic system, and is written in state space form as

$$\dot{\mathbf{x}} = \mathbf{A}_{lin} \cdot \mathbf{x} + \mathbf{B}_{lin} \cdot \mathbf{u}$$
$$\mathbf{y} = \mathbf{C}_{lin} \cdot \mathbf{x} + \mathbf{D}_{lin} \cdot \mathbf{u} \quad (15)$$

where the identified model is obtained through curve fitting [31] and has the parameters

$$\mathbf{A}_{lin} = \begin{bmatrix} -275.92 & -2.36\text{E}+05 & -6.43\text{E}+07 & -6.92\text{E}+08 \\ 1 & 0 & 0 & 0 \\ 0 & 1 & 0 & 0 \\ 0 & 0 & 1 & 0 \end{bmatrix}$$

$$\mathbf{B}_{lin} = \begin{bmatrix} 1 & 0 & 0 & 0 \end{bmatrix}^T \quad (16)$$

$$\mathbf{C}_{lin} = \begin{bmatrix} 0 & 0 & 0 & 6.90\text{E}+08 \\ 0 & 0 & 6.90\text{E}+08 & 0 \\ 0 & 6.90\text{E}+08 & 0 & 0 \end{bmatrix}$$

$$\mathbf{D}_{lin} = \begin{bmatrix} 0 & 0 & 0 & 0 \end{bmatrix}^T$$

and the resulting comparison of the transfer function is shown in Fig. 4.

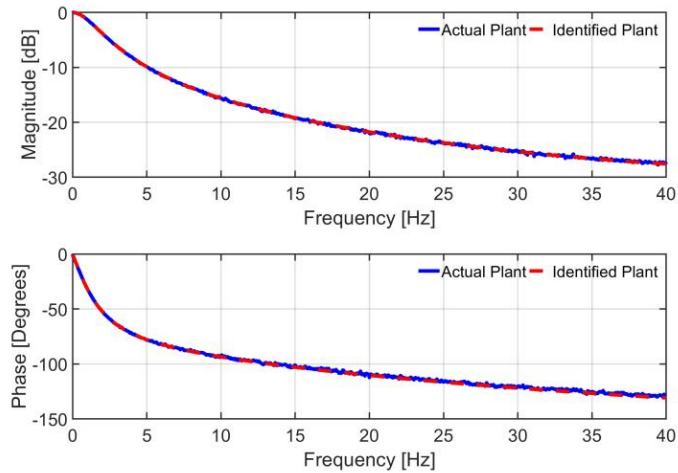

**Figure 4.** Identification process showing a representative comparison between the frequency responses of the *actual* plant and *linear* nominal plant from the plant command input (mm) to the actuator displacement (mm).

**Model Uncertainties**

To understand how the two approximate plant models (nonlinear nominal, and linear nominal) compare to the dynamics of the true plant, we evaluate the response of these three plant models using sinusoidal inputs at specific frequencies. Figure 5 shows the nonlinear relationship between the commanded displacement (input) and the measured displacement (output) for the three models. Note that for a 25 mm command input, the models seem to have fairly similar behavior to each other in general. The largest difference in the measured displacements occurs at the lower frequency range where the amplitude of the response is larger. Then, the dynamics of the actuator result in a gradually decreasing measured displacement as the frequency of the input increases.

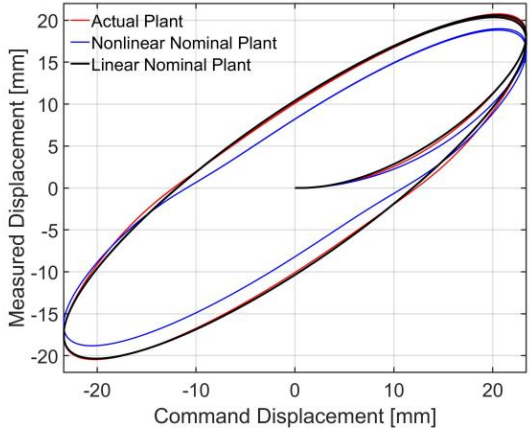

(a) Response at 1 Hz

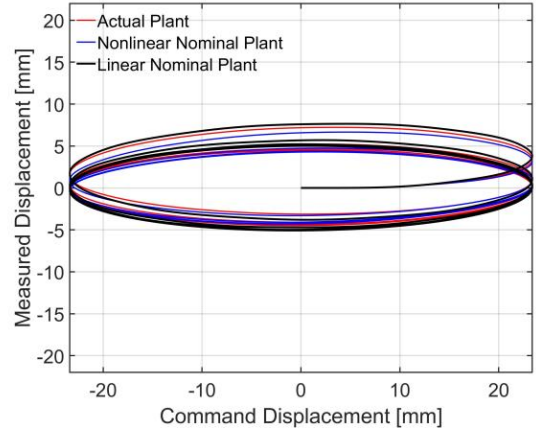

(b) Response at 8 Hz

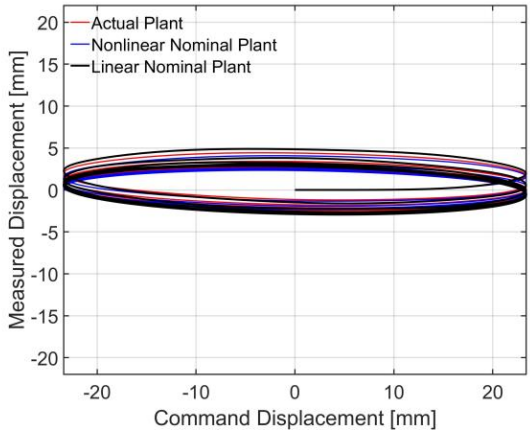

(c) Response at 14 Hz

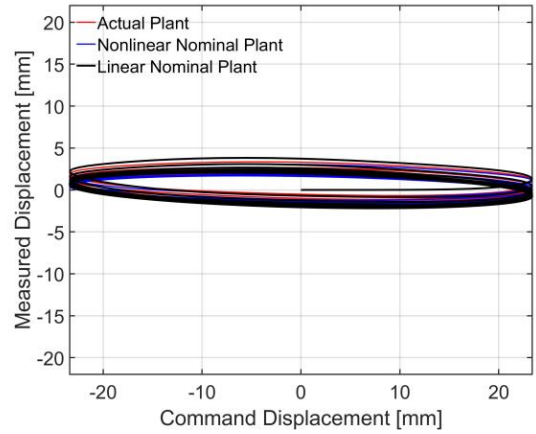

(d) Response at 19 Hz

**Figure 5.** Measured displacement vs. commanded displacement for the linear and nonlinear nominal plants and the actual plant.

To further evaluate the differences, the displacement response of these three plants is shown in the time-domain in Fig. 6 for three of the specific input frequencies used in Fig. 5. At 1 Hz the linear nominal model represents the behavior of the actual plant better than the nonlinear nominal plant. However, the nonlinear nominal model provides an improved approximation at higher frequencies where the response amplitude decreases. Quantitatively, taking the true plant as the reference, the normalized RMS error (NRMSE) of these two nominal models are computed using

$$NRMSE = \sqrt{\frac{\sum \left[ x_{nom}^{(n)}(i) - x^{(n)}(i) \right]^2}{\sum \left[ x^{(n)}(i) \right]^2}} \qquad (17)$$

where $x^{(n)}$ denotes the $n^{th}$ time derivative of the actual actuator displacement (the reference), and $x_{nom}^{(n)}$ denotes the corresponding $n^{th}$ time derivative of the displacement of each nominal model.

Each nominal model is quantitatively evaluated in terms of its deviation from the actual model. The results are used to compute the NRMSE values for each nominal model provided in Table 4. The linear model offers a description with low errors in displacement response. Only a 2% error in displacement is obtained for low frequencies, but the errors increase by an order of magnitude at higher frequencies. The nonlinear model provides a plant representation with the same level of error (~7 % for displacement) across the entire frequency range. The larger difference is observed in the time derivatives. For higher order states such as velocity and acceleration, the differences between the two nominal models are much more noticeable. The errors with the linear nominal model increase for the velocity and acceleration, while those for the nonlinear nominal model remain at the same level.

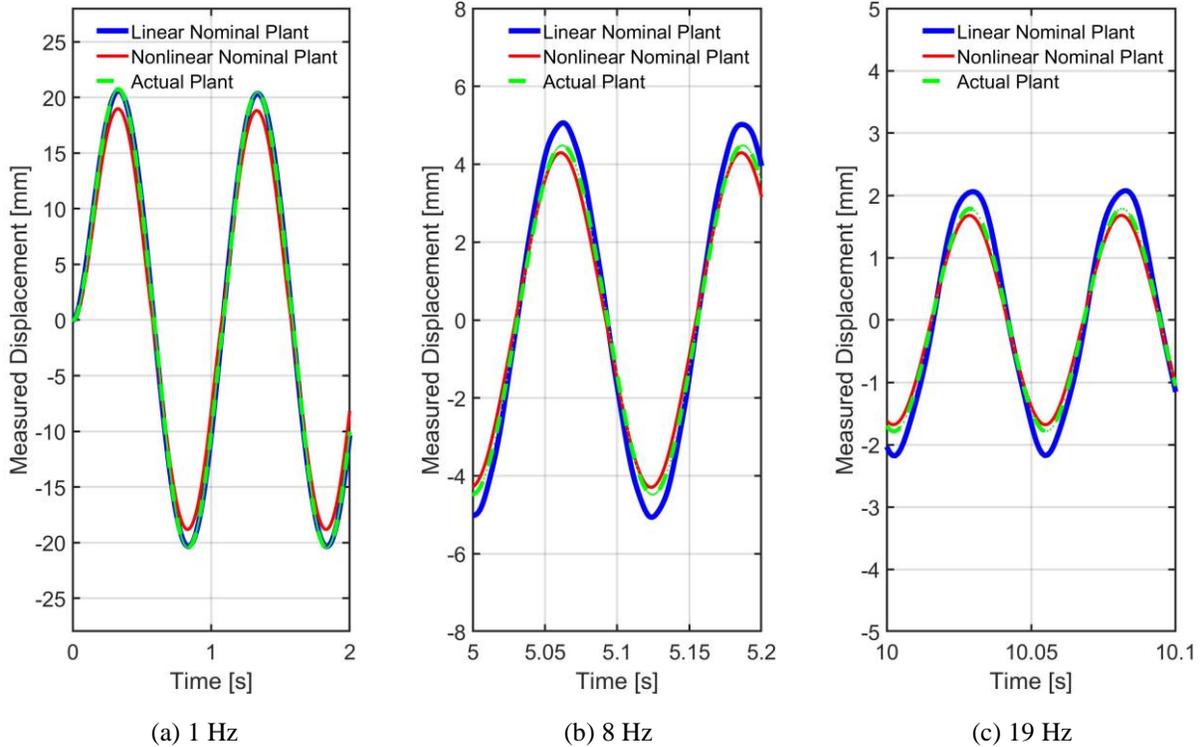

(a) 1 Hz    (b) 8 Hz    (c) 19 Hz

**Figure 6**. Time-domain displacement response of the actual plant and both nominal (linear and nonlinear) plants for a sinusoidal input at different frequencies.

**Table 4**. Normalized RMS error (%) between each nominal plant response with respect to the actual plant response

| Frequency Hz | Normalized RMS error (%) | | | | | |
|---|---|---|---|---|---|---|
| | **Displacement** | | **Velocity** | | **Acceleration** | |
| | Nonlinear nominal plant | Linear nominal plant | Nonlinear nominal plant | Linear nominal plant | Nonlinear nominal plant | Linear nominal plant |
| 1 | 10.1 | **1.97** | 10.8 | **6.85** | 20.6 | 94.9 |
| 8 | 7.50 | 14.0 | 7.73 | 15.6 | 10.1 | 57.4 |
| 14 | 6.90 | 17.9 | 6.89 | 20.8 | 7.91 | 60.6 |
| 19 | 6.91 | 19.9 | 6.86 | 24.7 | 7.69 | 63.8 |

## 3.1. State Estimation: Kalman filter and particle filter implementation

To study the capabilities of the particle filter for nonlinear estimation under uncertainty, we implement both estimators and compare the results. The Kalman filter is implemented based on the linear state-space model of the plant (Eq. 15) using a ratio $Q/R=1/100$, which was found to provide good state estimates. The optimal filter gain matrix is computed using the MATLAB function kalman.m.

Implementation of the particle filter requires that the dynamics of the nominal and actual plants are expressed in terms of the process and observation models (Eqs. 7 and 8, respectively) of the estimation algorithm. From this point on, all of the dynamic equations are written and implemented in discrete-time form. Table 5 shows the corresponding expressions. The controllable canonical form of the equation for the nonlinear nominal plant is the process model. Likewise, the controllable canonical form of the equation for the actual plant is the observation equation. These equations are integrated using a fifth-order Runge-Kutta algorithm [32,33] to compute the predicted states $\mathbf{x}_{k+1}$.

**Table 5**. Relationship between generalized and applied equations.

| Estimation model | System Transition Equation | Measurement Equation |
|---|---|---|
| Generalized problem | Eq. 7: $\mathbf{x}_{k+1} = f_k(\mathbf{x}_k, \mathbf{w}_k)$<br>*Process Model* | Eq. 8: $\mathbf{y}_k = h_k(\mathbf{x}_k, \mathbf{v}_k)$<br>*Observation Equation* |
| Plant dynamics | Eqs. 9 and 14:<br>*Nonlinear Nominal Plant* | Eqs. 9 and 13:<br>*Actual Plant* |

The process noise and measurement noise are considered here to be additive with known distributions. Thus, Eqs. 7 and 8 are written as

$$\mathbf{x}_{k+1} = f_k(\mathbf{x}_k) + \mathbf{w}_k \tag{18}$$

$$\mathbf{y}_k = h_k(\mathbf{x}_k) + \mathbf{v}_k \tag{19}$$

where the assumed distribution of the process noise is $\mathbf{w}_k \sim N(0, Q)$, and the assumed distribution of the measurement noise is $\mathbf{v}_k \sim N(0, R)$. The variance of the measured displacement and the measured force are $\sigma_d^2 = 1.07 \times 10^{-6} \, \text{m}^2$ and $\sigma_F^2 = 3.30 \times 10^3 \, \text{N}^2$, respectively. The variance of the process noise is set to $\sigma_P^2 = 2.67 \times 10^{-9} \, \text{m}^2$ and $\sigma_P^2 = 1.07 \times 10^{-9} \, \text{m}^2$ corresponding to a ratio $Q/R$ of 1/400 and 1/1000, respetively, which are used in the case of low and high noise levels, respectively.

The design of a particle filter uses the nonlinear nominal plant to move the particles from the previous time step to the current time step (see the prediction stage, Step 2 in Table 2). Figure 7 shows the essential difference regarding the nominal model used in each approach.

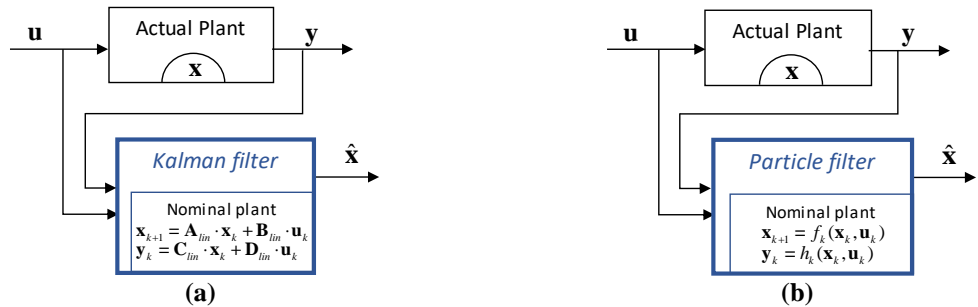

**Figure 7**. Estimation with the Kalman filter and the particle filter. Notice that the Kalman filter uses the linear nominal model (Eq. 15) and the particle filter uses the nonlinear nominal model (Eqs. 9 and 14) of the plant.

## 4. Results and Conclusions

To evaluate the ability of the two estimators to estimate the states of the actual nonlinear plant when modeling errors are present, they are subjected to a chirp input signal with the parameters provided in Table 6. A command input amplitude of 23.4 mm is chosen, corresponding to a highly nonlinear response of the actual plant. A chirp input is used because it has a broad frequency range. This choice is intended to meet the objective of investigating the dynamics of the estimators at different frequencies. Figure 8 shows the response of the actual plant to this input signal.

**Table 6.** Input signal: Chirp parameters.

| Parameter | Value | Unit |
|---|---|---|
| Initial Frequency | 0.1 | Hz |
| Final frequency | 20 | Hz |
| Amplitude | 23.4 | mm |
| Duration | 30 | sec |
| Sampling Frequency | 1024 | Hz |

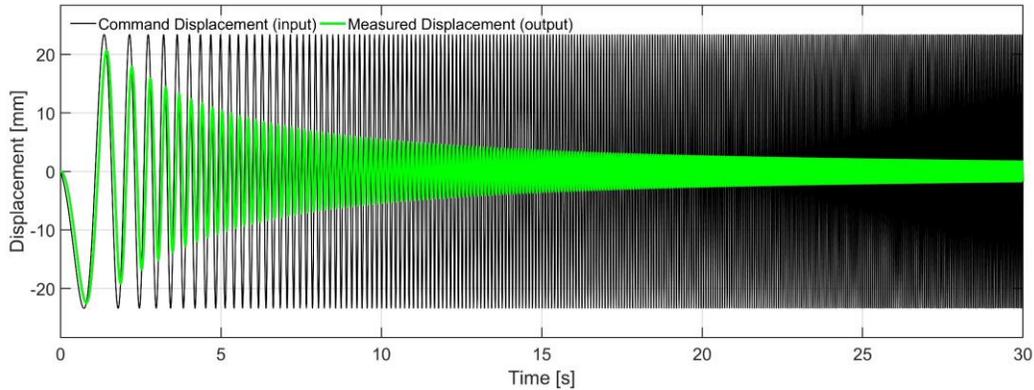

**Figure 8**. Displacement response of the actual plant to chirp input.

To systematically assess the effectiveness of the nonlinear estimator (particle filter), the amplitude is chosen to ensure the physical device exhibits strongly nonlinear behavior. The linear estimator with the Kalman filter uses the linear nominal plant to estimate the states and responses, while the particle filter uses the nonlinear nominal plant. In addition to the modeling errors present in the nominal models used to design and implement each estimator, three different levels of measurement noise are also included as sources of uncertainty, in which their amplitudes are defined as a percentage of the maximum displacement response of the plant. See Table 7 for a description of these noises. In this study we also evaluate the impact of the number of particles used on the ability of the nonlinear estimator to reach an acceptable estimation performance. Three values are used for the number of particles.

**Table 7.** Levels of measurement noise considered.

| Descriptive Term | Amplitude | Standard Deviation (mm) |
|---|---|---|
| Noise level 1 | 1 % | 0.2 |
| Noise level 2 | 5 % | 1.0 |
| Noise level 3 | 10 % | 2.1 |

The time-domain responses of the estimated states obtained with the two estimators are shown in Fig. 9, 10, and 11. The true and estimated responses are displayed at specific instances in time during the chirp to illustrate the performance variations with frequency. The actual plant model is used to obtain the displacement, velocity, acceleration, and force, and these are considered as the reference signals for comparison with the two estimation approaches. In these time-domain plots, the particle filter estimations are computed using 500 particles. As a quantitative measure of performance, NRMSE values are computed using the response within each of three consecutive time intervals (denoted case a: 0 – 10 sec, case b: 10 – 20 sec, and case c: 20 – 30 sec). These times correspond to three frequency ranges, denoted low, medium, and high, for assessing the performance of the estimators. The results of 20 simulations are averaged for each evaluation, and the equation for computing the NRMSE is

$$NRMSE = \sqrt{\frac{\sum \left[ \hat{x}^{(n)}(i) - x^{(n)}(i) \right]^2}{\sum \left[ x^{(n)}(i) \right]^2}} \tag{20}$$

where $x^{(n)}$ denotes the $n^{\text{th}}$ time derivative of the true actuator displacement, and $\hat{x}^{(n)}$ denotes its corresponding estimate.

Figure 9 illustrates several representative responses showing the actuator displacements obtained with each estimator at noise level 2. Here, for frequencies less than ~2.5 Hz (see Fig. 9a), both estimators perform similarly. However, at all other frequency ranges (Figs. 9(b-d)) the particle filter yields improved estimates. Furthermore, the performance gains are more significant as the frequency increases, which highlight the power of the particle filter to capture higher dynamics.

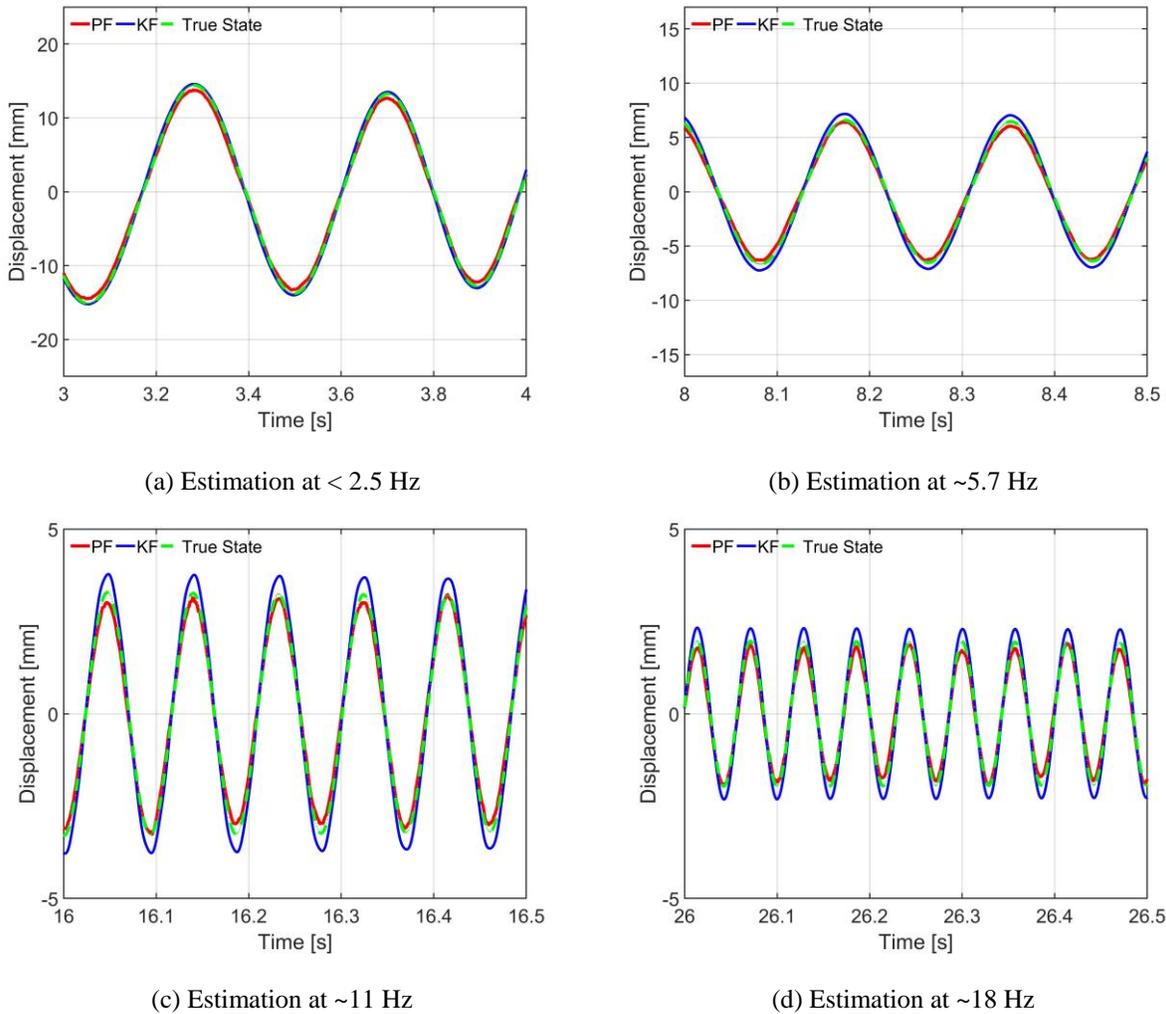

(a) Estimation at < 2.5 Hz
(b) Estimation at ~5.7 Hz
(c) Estimation at ~11 Hz
(d) Estimation at ~18 Hz

**Figure 9**. Estimated displacement comparison: Kalman filter (KF) and particle filter (PF) with 500 particles.

Figure 10 provides a comparison of the performance of the estimators when the actuator acceleration is estimated. Here the particle filter is able to capture the plant dynamics at all frequency ranges considered, while the Kalman filter exhibits high frequency dynamics that are not present in the true system. The particle filter does not generate such undesirable higher frequency dynamics.

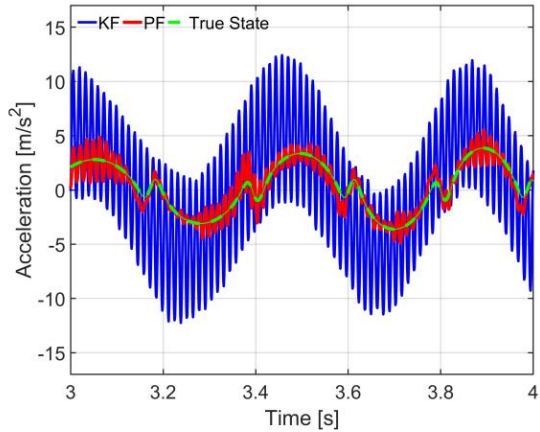

(a) Estimation at < 2.5 Hz

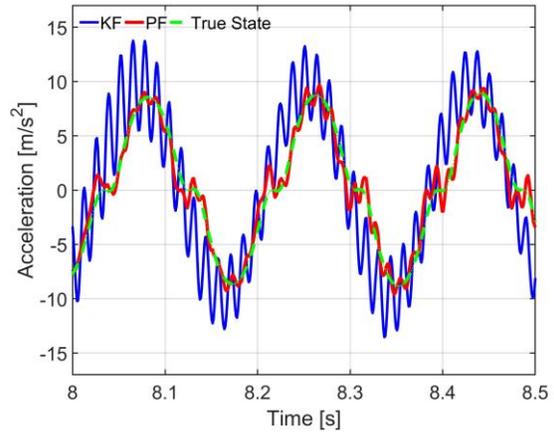

(b) Estimation at ~5.7 Hz

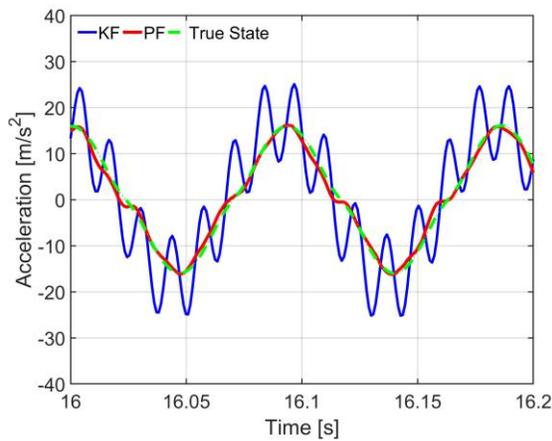

(c) Estimation at ~11 Hz

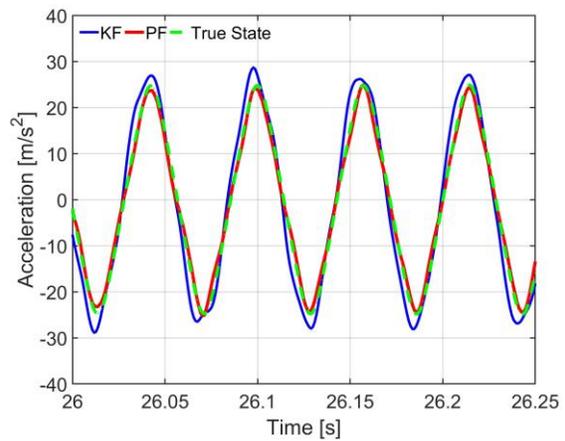

(d) Estimation at high frequency (~18 Hz)

**Figure 10**. Estimated acceleration comparison: Kalman filter (KF) and particle filter (PF) with 500 particles.

Figure 11 shows the performance of the particle filter when estimating the actuator force. In this case, the measured force is also shown, with the measurement noise included, to highlight the capabilities of the nonlinear estimator. The nonlinear estimator produces responses that are much closer to the true state at lower frequencies, revealing the increasing difficulty of estimating forces as the frequency increases.

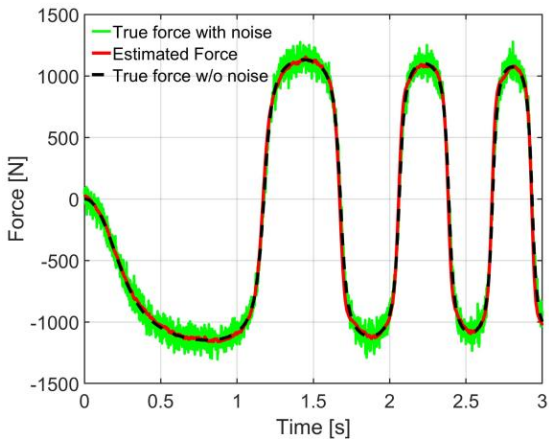
(a) Estimation at < 2.5 Hz

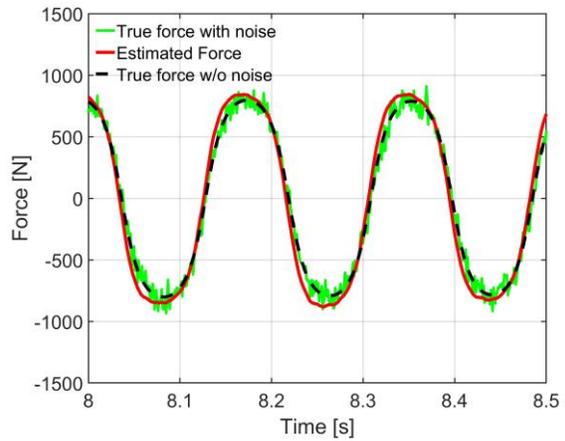
(b) Estimation at ~5.7 Hz

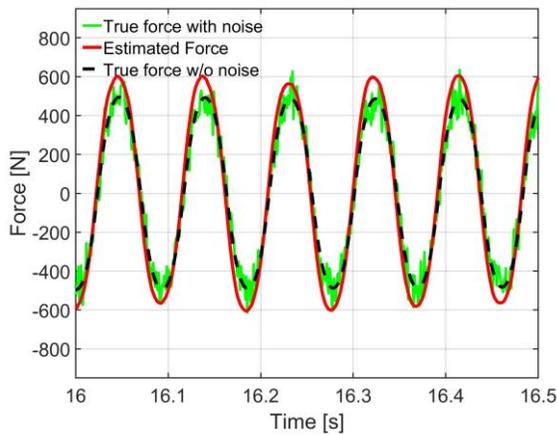
(c) Estimation at ~11 Hz

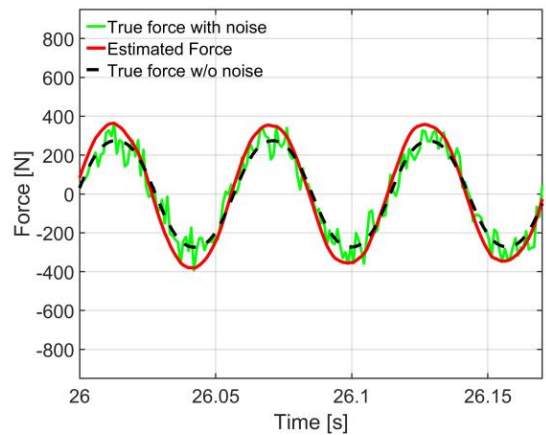
(d) Estimation at ~18 Hz

**Figure 11**. Estimated force: Particle filter (PF) estimation with 500 particles.

Figure 12 provides quantitative results for the cases presented (at noise level 2) and describes the influence of the number of particles used. Twenty realizations are averaged to obtain the NRMSE values in each case. For the uncertainties considered here, shown in Table 4, the particle filter estimator needs ~100 particles to outperform the Kalman filter results: the estimation errors of displacement, velocity, and acceleration for particle filter are around 50%, 40%, and 30% of the Kalman filter errors, respectively, except for the displacement at very low frequencies of ~ 0.2 Hz, where 500 particles are required to provide comparable NRMSE values. This requirement is explained by the fact that the displacement response of the nonlinear nominal model has a difference of ~10% at low frequencies with respect to the actual plant (see Table 4), versus a 2% corresponding error in the linear nominal model. Thus, to compute an improved estimate of the displacement, the posterior probability distribution of the state needs to be enhanced by increasing the number of particles. As expected, increasing the number of particles generally decreases the NRMSE values.

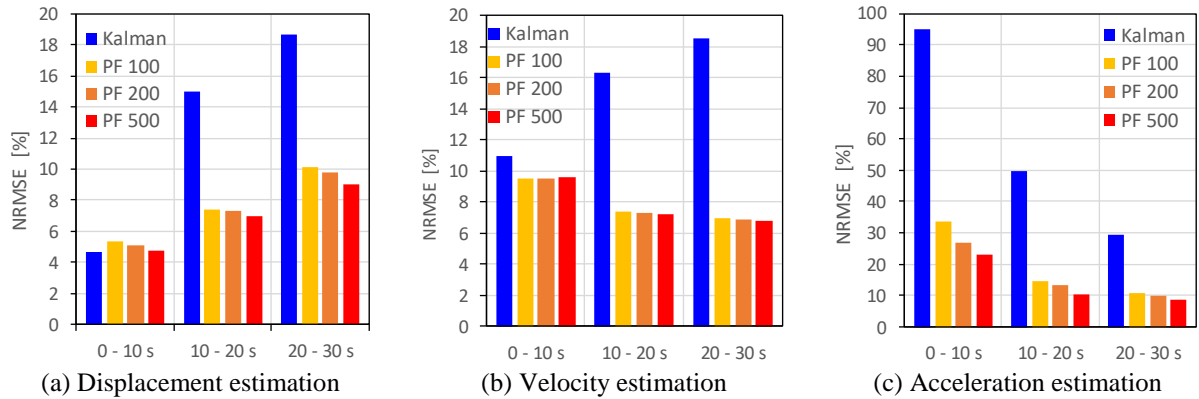

(a) Displacement estimation (b) Velocity estimation (c) Acceleration estimation

**Figure 12**. Influence of the number of particles (100, 200, and 500) utilized in estimating the` states for noise level 2. Measures shown are normalized RMS errors.

Figure 13 provides the NRMSE values for various noise cases presented previously (see Table 7). The particle filter results shown are computed using 500 particles. Twenty realizations are averaged to obtain each of these. The particle filter typically yields NRMSE values that are much less than those obtained with the Kalman filter. Furthermore, the Kalman filter has a tendency to yield larger errors with increasing frequencies even with low levels of noise. In contrast, the particle filter is able to maintain low errors in the estimates with increasing frequency, except for large levels of noise at a high frequency range. In the particle filter, this ability can be effectively controlled by increasing the ratio $Q/R$ (making the estimator trust the nominal model more than in the measurement for high level of noise, see Section 3.1) yielding a decreasing trend in NRMSE with frequency for all noise cases. Conversely, increasing the ratio $Q/R$ in the Kalman design does not provide an important reduction in the NMRSE measures.

Similar conclusions are evident for the states corresponding to the higher-order time derivatives. Figures 14 and 15 show the NRMSE measures of performance for velocity and acceleration estimation, respectively. Note that the measurement noise heavily influences the results of the Kalman filter. If the estimated states from the Kalman filter were used as feedback to a controller, these may lead to instabilities. On the contrary, the particle filter exhibits a decreasing trend in NRMSE with frequency, and is able to both deal with modeling errors and measurement noise. Generally, the particle filter yields a more stable estimate of the states with both classes of uncertainty.

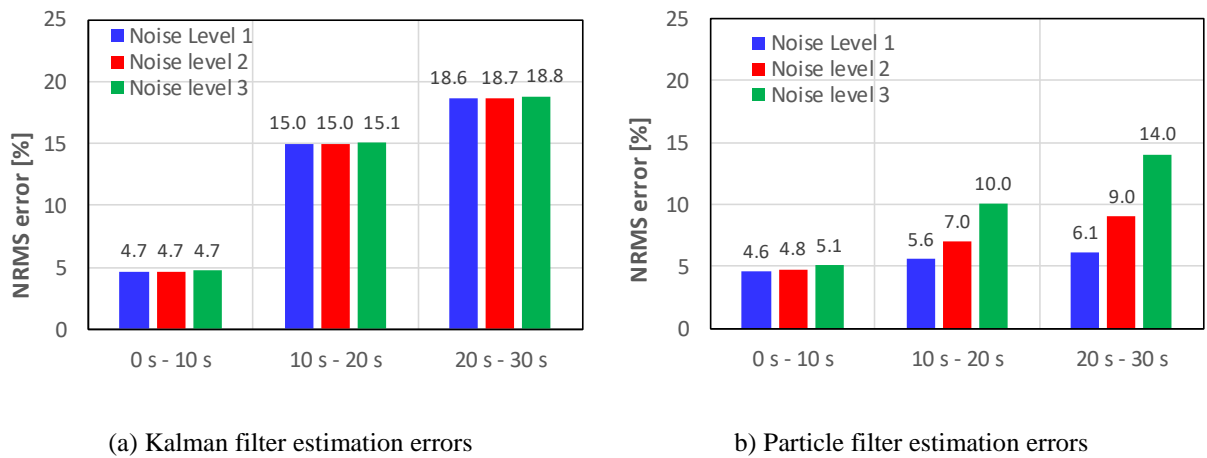

(a) Kalman filter estimation errors  b) Particle filter estimation errors

**Figure 13**. NRMSE values for displacement estimates for specific intervals of time.

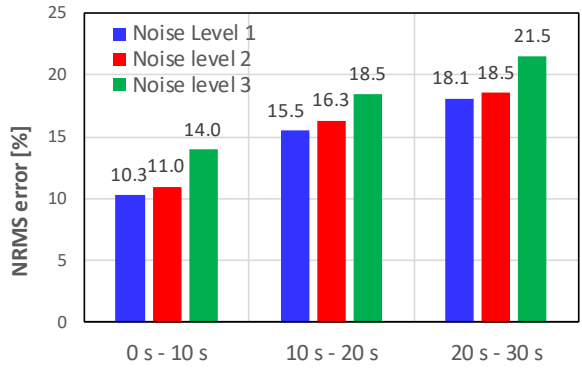
(a) Kalman filter estimation errors

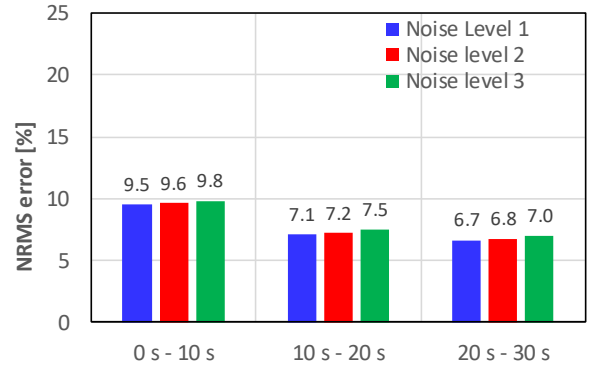
b) Particle filter estimation errors

**Figure 14**. NRMSE values for velocity estimates for specific intervals of time.

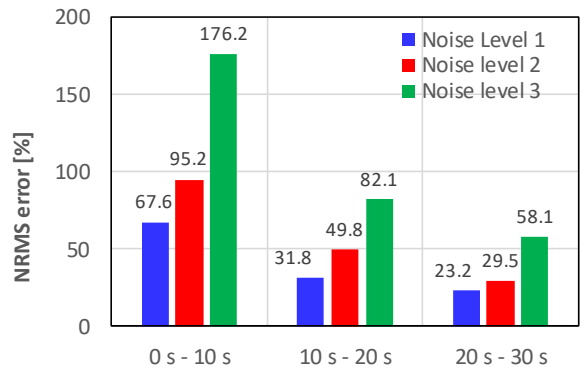
(a) Kalman filter estimation errors

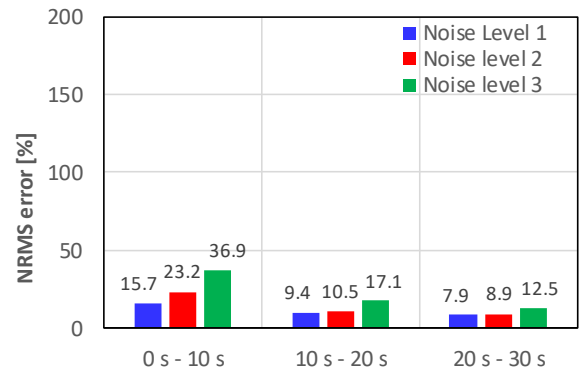
b) Particle filter estimation errors

**Figure 15**. NRMSE values for acceleration estimates for specific intervals of time.

Another aspect worth investigating for real-time applications is the variation of the estimated states for different realizations. This analysis is to examine the statistics of the NRMSE values for the estimated states for the three-time intervals defined previously with noise level 3. As before, running both approaches 20 times and using the particle filter with the smallest number of particles considered in this study, the mean and standard deviation of the NRMSE measures are computed. Table 8 shows the first two statistical moments for the displacement and its higher order derivative states. For higher order states, the particle filter demonstrates its superiority by yielding smaller standard deviations. The average standard deviation with the Kalman filter is close to 10 % of the mean for the velocity and 25% for the acceleration, whereas the particle filter average standard deviation is about 1% and 5% for the velocity and acceleration NRMSE, respectively. In the case of the displacement, the particle filter exhibits a slightly larger standard deviation, 1.4% compared to 1% for the Kalman filter. Although this is a reasonable value, it can be improved easily by increasing the number of particles. In this analysis, the objective is to demonstrate that the particle filter provides precise estimates with a small variation among realizations if we see a set of simulations as an ensemble of realizations.

Table 8. Mean and Standard Deviation of the NRMSE values for noise level 3.

| Estimated State | Time interval | Mean | | Standard Deviation | |
|---|---|---|---|---|---|
| | | Kalman filter | Particle filter (100 particles) | Kalman filter | Particle filter (100 particles) |
| Displacement | 0 - 10 s | 4.71 | 5.52 | 0.01 | 0.14 |
| | 10 - 20 s | 15.05 | 10.44 | 0.03 | 0.28 |
| | 20 - 30 s | 18.78 | 15.91 | 0.11 | 0.43 |
| Velocity | 0 - 10 s | 14.55 | 10.06 | 1.67 | 0.14 |
| | 10 - 20 s | 18.38 | 7.91 | 1.35 | 0.08 |
| | 20 - 30 s | 20.99 | 7.49 | 1.93 | 0.07 |
| Acceleration | 0 - 10 s | 188.31 | 55.70 | 38.44 | 3.33 |
| | 10 - 20 s | 79.46 | 25.09 | 17.61 | 1.31 |
| | 20 - 30 s | 52.69 | 16.57 | 14.79 | 0.61 |

# Conclusions

Nonlinear dynamics present a particular challenge in dynamic testing involving both estimation and control. This study investigates the performance of a particle filter for use as a nonlinear estimator of the hidden states of a nonlinear system in the presence of uncertainties. Considering a plant consisting of a servo-hydraulic actuator and a nonlinear physical specimen, representing a typical setup for a real-time hybrid simulation experiment, the performance of the particle filter for nonlinear state estimation is directly compared to a linear Kalman filter estimator in a numerical simulation. Uncertainties are included in the form of both modeling errors and measurement noise. In these numerical simulations, both time history estimates and quantitative measures consisting of normalized RMS errors are compared to study the performance gains possible with the particle filter. The particle filter is adopted and implemented to design a recursive nonlinear estimator. The particle filter is shown to yield significantly better performance in estimating the displacement, velocity, acceleration and force of the plant. Over a range of frequencies, and for various noise levels, the particle filter consistently yields lower quantitative error values. The results indicate that this Bayesian approach provides accurate estimates of states over a broad frequency range, while maintaining its effectiveness in the presence of measurement noise. Accurate estimates of the higher-order derivatives of the states are obtained, showing the superiority of this approach compared to estimators used currently in practice. This method has notable advantages for used in advanced real-time control techniques such as real-time hybrid simulation.

## Acknowledgement

This work was supported by Purdue University through the John E. Goldberg Fellowship and through the Peruvian National Council of Science, Technology, and Technological Innovation (CONCYTEC) Fellowship *Generación Científica: Becas de Doctorado en el Extranjero*.